# Hybrid Approach to Identify Druglikeness Leading Compounds against COVID-19 3CL Protease

Imra Aqeel[1,2], Abdul Majid[1,2]

[1]Biomedical Informatics Research Lab, Department of Computer & Information Sciences, Pakistan Institute of Engineering & Applied Sciences, Nilore, Islamabad 45650, Pakistan

[2]PIEAS Artificial Intelligence Center (PAIC), Pakistan Institute of Engineering & Applied Sciences, Nilore, Islamabad 45650, Pakistan

imraaqeel@pieas.edu.pk

*Abstract—* SARS-COV-2 is a positive single-strand RNA-based macromolecule that has caused the death of more than 6.3 million people since June 2022. Moreover, by disturbing global supply chains through lockdown, the virus has indirectly caused devastating damage to the global economy. It is vital to design and develop drugs for this virus and its various variants. In this paper, we developed an in-silico study-based hybrid framework to repurpose existing therapeutic agents in finding drug-like bioactive molecules that would cure Covid-19. We employed the Lipinski rules on the retrieved molecules from the ChEMBL database and found 133 drug-likeness bioactive molecules against SARS coronavirus 3CL Protease. Based on standard IC50, the dataset was divided into three classes active, inactive, and intermediate. Our comparative analysis demonstrated that the proposed Extra Tree Regressor (ETR) based QSAR model has improved prediction results related to the bioactivity of chemical compounds as compared to Gradient Boosting, XGBoost, Support Vector, Decision Tree, and Random Forest based regressor models. ADMET analysis is carried out to identify thirteen bioactive molecules with ChEMBL IDs 187460, 190743, 222234, 222628, 222735, 222769, 222840, 222893, 225515, 358279, 363535, 365134 and 426898. These molecules are highly suitable drug candidates for SARS-COV-2 3CL Protease. In the next step, the efficacy of bioactive molecules is computed in terms of binding affinity using molecular docking and then shortlisted six bioactive molecules with ChEMBL IDs 187460, 222769, 225515, 358279, 363535, and 365134. These molecules can be suitable drug candidates for SARS-COV-2. It is anticipated that the pharmacologist/drug manufacturer would further investigate these six molecules to find suitable drug candidates for SARS-COV-2. They can adopt these promising compounds for their downstream drug development stages.

## 1. Introduction

Novel coronavirus (nCoV-19) is a rapidly spreading pandemic. International Committee on Taxonomy of Viruses (ICTV) officially named severe acute respiratory syndrome coronavirus 2 (SARS-CoV-2) on February 11, 2020 [1]. The first time, coronavirus appeared in December 2019 from Asia and then spread out worldwide. A total of 228 countries and more than 500 million people got infected. SARS-CoV-2 is alike MERS-CoV and SARS-CoV. Both these viruses have caused severe acute respiratory syndrome. There are seven strains of Alpha and Beta coronaviruses in human coronaviruses. HCoV-229E and HCoV-NL63 belong to the type of alpha-coronaviruses. On the other hand, HCoV-HKU1, HCoV-OC43, SARS-CoV, MERS-CoV, and SARS-CoV-2 belong to beta-coronaviruses [2]. COVID-19 virus is a single-strand ribonucleic acid (ssRNA) virus that consists of multiple structural and non-structural proteins. The structural proteins have four different types spike (S), membrane (M), envelope (E), and nucleocapsid (N) proteins. However, non-structural proteins contain sixteen different types named NSP1, NSP2, NSP3…, and NSP16. These proteins are mainly more responsible for spreading out SARS-CoV-2 than other types of proteins. Consequently, these proteins are considered potential targets to prevent SARS-CoV-2, especially the 3C-like protease (3CL$^{pro}$ or M$^{pro}$) is crucial for replication [3]. Fig.1.a shows a visual model of SARS-COV-2 with all constituent proteins and Fig.1.b depicts its large genome size of 29.9 kb starting from 5´ to 3´. This virus has the inherent capability of auto-reproduction into sixteen different types of non-structural proteins.

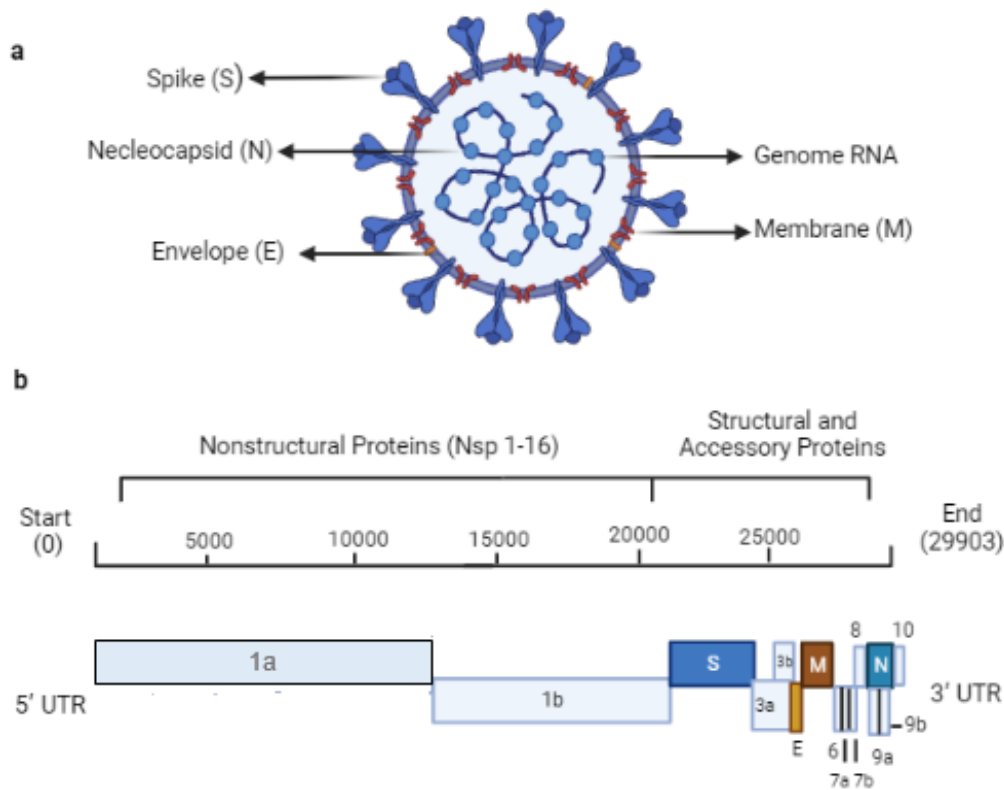

Fig.1. (a) SARS-COV-2 with constituent proteins, (b) Related genome detailed information

Upon entrance into the host cell, the viral genome is translated to produce two overlying polyproteins named *pp1a* and *pp1b* [4]. During the proteolytic activity, these polyproteins are excised from the 3CL protease (3CL$^{pro}$, also known as Main protease (M$^{pro}$). These proteins work with papain-like protease to slice the polyproteins to produce a total of sixteen functional nonstructural proteins (NSPs). It was reported that the eleven slicing sites of polyprotein 1ab were shared and operated by only the 3CL$^{pro}$ of SARS and no other human protease involve in slicing processing [4]. To initiate viral replication, the viral replication transcription complex (RTC) is assembled by the sliced NSPs.

The computational drug discovery process has become a crucial strategy to develop the drug against COVID-19. It can be an effective tool to save money and reduce the time for drug discovery/ repurposing [5]. Recent, machine learning (ML) approaches are employing advanced data mining and data analytics techniques. The online public available medical databases contain sufficient information related to bioactive molecules. This has made it possible to develop ML approaches based QSAR model to quickly develop vaccines for the COVID-19 pandemic [6]. Due to stringent storage requirements, this vaccine is rather difficult to transport and warehouse. Moreover, successful virus vaccinations for humans and animals are seriously hampered by vaccine-associated increased illness [7]. It has shown that people are not as receptive to getting vaccinated as they are to taking drugs [8]. On the other hand, underdeveloped countries suffered the most from the pandemic with the official death tolls of India and Brazil, at the time of writing this manuscript, being 525,000 and 672,000, respectively. Since the start of this pandemic, in India and Indonesia, millions of people have been pushed back into poverty [8]. Such nations would benefit from cheap, easy-to-store, and rapidly deployable effective drugs against SARS-COV-2.

In this study, we repurposed existing therapeutic agents by examining drug-like bioactive molecules for Covid-19. For this purpose, we have developed a hybrid approach that combines the useful extracted information using bioinformatics tools such as the SWISSADME platform, AUTODOCK VINA software, and PYMOL. The cheminformatics information related to molecular descriptors is obtained using the RDKIT tool. For QSAR modeling, we have selected six diverse types of regression algorithms such as Extra Tree, Gradient Boosting, XGBoost, Support Vector, Decision Tree, and Random Forest. These regression algorithms are most commonly used in the literature on cheminformatics for drug discovery. Our comparative analysis demonstrated that Extra Tree Regressor (ETR) based QSAR model has improved prediction results related to the bioactivity of chemical compounds as compared to other ML-based QSAR models. We used Lipinski rules on the retrieved molecules from the ChEMBL database and found 133 drug-likeness bioactive molecules against SARS coronavirus 3CL Protease. From ADMET analysis, we have identified thirteen novel bioactive molecules for SARS-Cov-2. In the next step, the efficacy of bioactive molecules is computed in terms of binding affinity and short-listed six bioactive molecules with ChEMBL IDs 187460, 222769, 225515, 358279, 363535, and 365134.

The rest of the paper is arranged as follows: Literature review of recent methods used to identify lead compounds is described in section 2. In section 3 material and proposed hybrid framework is explained. Results and discussions are presented in section 4. Section 5 highlights the conclusion of this study.

## 2. Literature Review

In recent years, machine learning algorithms are being used in the development of vaccines and drug development processes. In particular, several efforts have been carried out to perform virtual screening of bioactive molecules that could inhibit SARS-Cov-2. In a study [9], authors developed a unique drug-similarity model using the characteristics of existing drugs like remdesivir, dexamethasone, and baricitinib to inhibit COVID-19. They retrieved the interactive compounds using the known chemical-chemical interaction to repurpose existing drugs against SARS-CoV-2. They developed a two-tier clustering approach in which tier-1 used the t-Distributed Stochastic Neighbor Embedding (t-SNE) and tier-2 analyzed the two cluster analysis. Then molecular docking was performed to check the validation of top drug candidates. In research [10], authors developed a network-based model to explore the drug candidates against COVID-19. They used the genome similarity among SARS-COV-2 and other viruses like SARS and MERS. They designed a molecular network and found 30 suitable drugs including chloroquine, thalidomide, and rorgrapholide. In another work [11], the authors performed virtual screening to repurpose the drugs against COVID-19. They identified the two existing drugs as Lurasidone and Talampicillin. They also identified two drug-likeness molecules from the Zinc database. In the work, they performed molecular dynamic simulation and ADMET analysis was also carried out.

In another research [12], the authors proposed a deep learning approach using SVM, logistic regression, and Random Forest to calculate the molecular descriptor. From QSAR modeling, they calculated the binding affinities of proteins with the drug target. In [13], the authors performed quantitative high-throughput screening (qHTS) to investigate the potential inhibitors against SARS-COV-2 3CL Protease. In [2], the authors explained that natural bioactive compounds especially from plants can be the potential inhibitors against SARS-COV-2. They presented the viral biology properties and incursion mechanism. They also discussed the role of various proteins such as ACE22, Spike, 3CL Protease, TMPRSS2, and helicase. In [14], the authors discussed the uses and limitations of bioinformatics tools to prevent and reduce the spread of SARS-COV-2. In [15], the authors described that niacin would be the potential therapy for COVID-19. They explored the properties of CRC patients and investigated the prognosis, biological functions, survival rate, and binding capacity. In [16], the authors proposed some FDA-approved drug candidates for the therapeutic of COVID-19. They used Jaccard similarity analysis on the lung cancer drugs dataset taken from DrugBank and PubChem using graph neural network models.

In [17], the authors used the method of virtual screening and found multiple drug candidates. After performing further analysis two antiviral candidates and bafetinib and 7-hydroxystaurosporine showed better results. In [18], the authors described the various machine learning approaches used for Drug Target Interaction and discussed their advantages and limitations. In [5], the authors discussed the various methods of molecular docking, Disease-Disease Interaction, network model, MD simulation, AI, and machine learning models to design the Anti-COVID-19 drugs. In [19], the authors proposed the shape similarity-based pre-docking and interaction similarity-based post-docking methods to screen the drugs against COVID-19.

In [20], the authors analyzed the literature-based discovery methods to reposition the drugs for COVID-19. They also compared the three literature-based methods named BITOLA, Arrowsmith, and SemBT. In [21], the authors suggested a graphical neural network based on learning of embedding of chemical compounds to predict molecular properties for COVID therapy. In [1], the authors discussed the biology of COVID-19, and also discussed the healing development including vaccines against COVID-19. In [22], the authors investigated the seven machine learning models and four deep learning methods to find the chemical compounds against COVID-19 through the Ligand Based Drug Designing approach. In [23], the authors used the Naïve Bayes Machine Learning Algorithm and Drug Bank to screen the anti-COVID compound.

In [24], the authors used the registered clinical trials to repurpose the drug against COVID-19. In [25], the authors developed a database named DockCov2 containing FDA-approved drugs to accelerate the research in drug repurposing for COVID-19. In [26], the authors used the scaffold of compounds containing N-N-C(S)-N with spike protein to train and test the ML-based RF approach. In [27], the authors proposed the ML model to compare the efficacy of antiviral drugs against COVID-19. In [28], the authors used a plaque reduction assay to investigate the efficacy of antiviral drugs against COVID-19. In [29], the authors performed the ligand-based strategy with KNIME analytics for two diseases GLUT-1 and COVID-19. In [30], the authors performed virtual screening on Drug Bank to investigate the antiviral drugs against COVID-19.

## 3. Materials and Methods

The proposed hybrid framework is divided into four modules as shown in Fig.2. This figure shows that module A has different steps involved in the preparation of the input dataset. However, module B described the development of the QSAR model and performance comparisons with different state-of-the-art ML models. Module C explains how ADMET analysis is carried out for bioactive molecules. Finally, Module D demonstrates how molecular docking is carried out to validate results obtained from ADMET analysis.

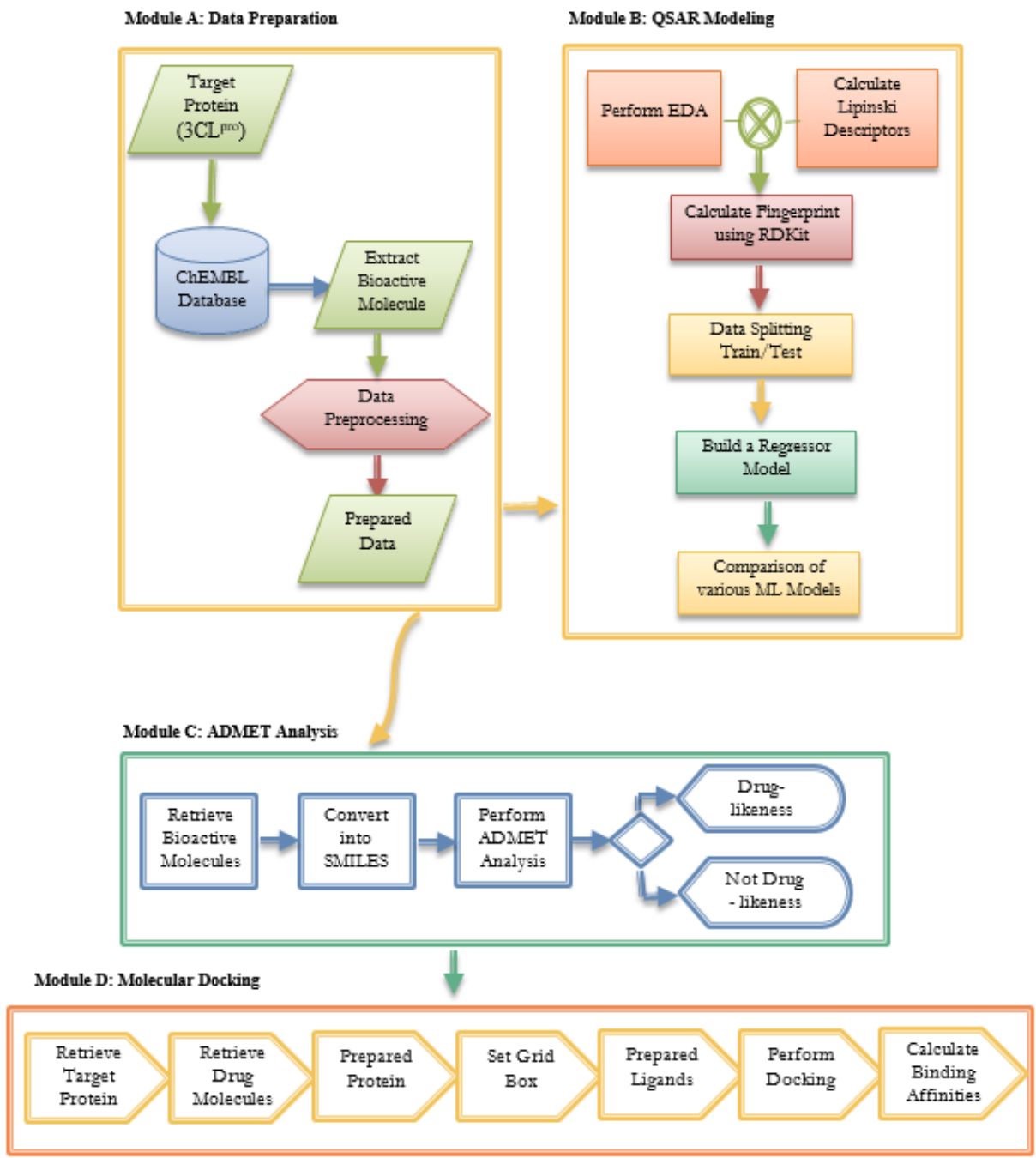

Fig.2. Demonstrate four main modules (A to D) in the proposed hybrid framework.

### 3.1. Module A: Data Preparation

Module A explains different data preprocessing stages as follows:

#### 3.1.1. Targeting Replicating Enzyme

Due to viral enzyme properties, the Main protease, 3C-like protease (3CL$^{pro}$) is a potential drug target among coronavirus proteins. This protease with papain-like protease (PL$^{pro}$) in the viral

RNA plays a pivotal role in replication and transcription [31]. It is also called Main Protease (M$^{pro}$). It is a highly conserved replicated key enzyme. Due to such crucial characteristics of 3C-like protease, it is an attractive potential target to investigate the binding inhibitors that can effectively bind the target protein.

### 3.1.2. Dataset

A dataset of inhibitors (Small Molecules) against SARS coronavirus 3C-like proteinase (Target ID: CHEMBL 3927) is retrieved from the ChEMBL database [32] that is publically available at https://www.ebi.ac.uk/chembl/. The ChEMBL database is manually abstracted from the published literature [32]. It contains bioactive molecules having drug-like features with combined chemical, functional, bioactivity, and genomic data to contribute to the transformation of genomic information into new potent drugs. It also contains information for the molecules related to the ADMET criteria. Currently, this database consists of 18.6M bioactivity measurements for more than 2.1 M compounds and 14K protein targets. Information from more than 81,000 research publications has been extracted to develop this database. Our retrieved dataset consists of 133 bioactive molecules out of 8.2 K compounds. These extracted small inhibitory molecules also contain the values of the standard type of IC50.

### 3.1.3. Data Preprocessing

The description related to the dataset as explained in the above section consists of 133 small inhibitory molecules. These bioactive molecules are measured in standard unit IC50 values in nM (nanoMol). The molecules with no IC50 values are dropped. Duplicated data is also deleted. To normalize the IC50 data distribution, we have taken each bioactive compound to its binding affinity to a target protein and converted it into pIC50 (pIC50 = -log10 (IC50)). Those bioactive molecules having low pIC50 values (< 04) are removed. After cleaning and preprocessing, the dataset consists of 86 small bioactive molecules. Next, bioactive compounds are labeled as either active, inactive, or intermediate classes based on their IC50 values. Active class compounds are those having IC50 values less than or equal to 1000 nM. However, inactive class compounds have IC50 values greater than or equal to 10000 nM. The compounds with IC50 values between 1001 nM to 9999 nM are labeled as an intermediate class.

## 3.2. Module B: QSAR Modeling

In our in-silico study, Quantitative structure-activity relationship (QSAR) models are developed to predict the chemical compounds having the best unknown biological activities. QSAR is a mathematical modeling method for predicting relationships between structural properties of known chemical compounds with their unknown biological activities. In QSAR modeling, each compound is characterized by its molecular descriptors and then the model can be used to predict how the change in structural property causes a change in biological activity [33]. Structural properties refer to physicochemical properties that represent the structure. However, biological activities refer to pharmacokinetic properties.

### 3.2.1. Exploratory Data Analysis (EDA)

To check the drug-likeness of the bioactive compounds, Lipinski descriptors are calculated. Lipinski, a scientist of Pfizer described a set of rule-of-thumb to evaluate the drug-likeness of a chemical compound. The rule outlines the molecular characteristics of a drug's pharmacokinetics—its absorption, distribution, metabolism, and excretion—that represent how well the drug works in the body ("ADME") [33]. Lipinski's rule describes that, in general, an orally active drug should not violate more than one condition of the following criteria:

- Molecular Weight (MW) should be less than 500 Dalton
- Octanol-water partition coefficient (LogP) should be less than 5
- Hydrogen-bond-donors (NumHDonors) should be less than 5
- Hydrogen-bond-acceptors (NumHAcceptors) should be less than 10

By examining Lipinski's rule-of-five descriptors, the chemical space of 3CL inhibitors was navigated to get insight into the structure-activity connection. This chemical space analysis may offer vital information about the fundamental characteristics of substances that control their inhibitory properties. Furthermore, exploratory data analyses via Lipinski descriptor are performed. Fig.3.a shows the frequency plot of active class and inactive class. However, Fig.3.b depicts a scatter chart comparing MW with LogP. This figure demonstrates that both bioactivity classes are straddling similar chemical regions.

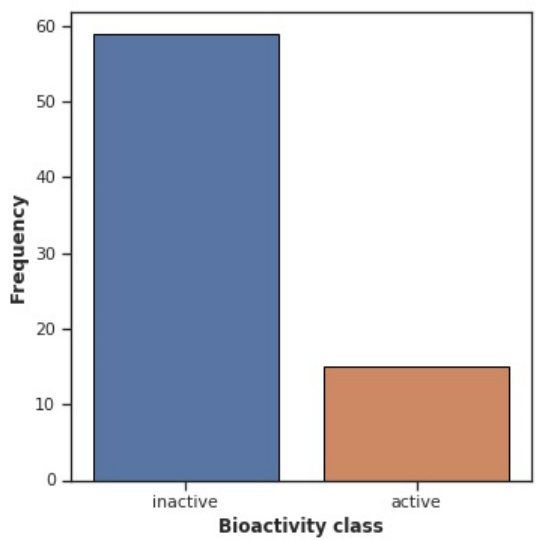
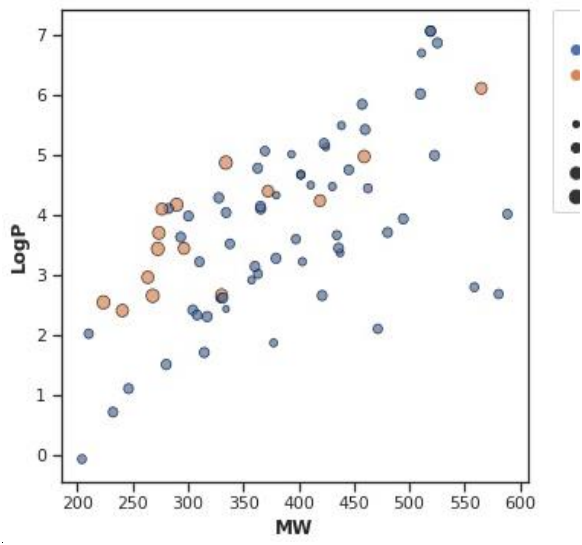

Fig.3.(a) Frequency plot of Bioactivity Classes    Fig.3.(b) Scatter plot of MW vs LogP

### 3.2.2. Feature Extraction

PubChem database [34] is used for feature extraction from inhibitory molecules. This is one of the largest databases that possess chemical structure and bioactive molecules. Our extracted dataset contained 86 small inhibitory molecules that are in SMILES format. The SMILES notations are converted into PubChem Substructure Fingerprints. PaDEL descriptor is used to calculate the PubChem Substructure Fingerprints. Before converting the SMILES into fingerprints, SMILES are cleaned from salt and standardizing tautomer as well, using built-in functions available in the PaDEL descriptor [33]. This feature set is used for training the proposed model. Table 1 illustrates the description of PubChem substructure fingerprints. It consists of 881 columns in the form of an order list of 0/1 bit: bit position 0-114 in PubChem Substructure Fingerprints represent the presence of chemical atoms that are represented by atomic symbols; bit position 115-262 represents the presence of the described chemical ring system; bit position 263-326 denotes the simple atom pairs; bit position 327-415 represents the simple atom nearest neighbors; bit position 416-459 represents detailed atom neighborhoods; bit position 460-712 imply the simple SMARTS patterns and bit position 713-880 signify the complex SMARTS patterns.

Table 1. PubChem Substructure Fingerprints description

| Bit Position | Description |
| --- | --- |
| 0-114 | Presence of chemical atoms |
| 115-262 | Presence of the described chemical ring system |
| 263-326 | Presence of simple atom pairs |
| 327-415 | Presence of simple atoms nearest neighbors |
| 416-459 | Presence of detailed atom neighborhoods |
| 460-712 | Presence of simple SMARTS patterns |
| 713-880 | Presence of complex SMARTS patterns |

### 3.2.3. Extra Tree Regressor-based Ensemble Model

The objective of this study is to build the regression models that enable the estimation of the continuous response variable (i.e., pIC50) being a function of predictors (i.e., PubChem fingerprint descriptors). For this purpose, various ML algorithms are developed for QSAR modeling. Due to higher prediction performance, we select the Extra Tree Regressor (ETR) based ensemble approach. This model employs a meta-estimator to fit several randomized decision trees and during training pick diverse dataset sub-samples. This algorithm avoids over-fitting and provides improved predictive accuracy.

Dataset explained, in section 3.1.3 consists of 86 bioactive molecules, and is used to train and test the model. Before using this dataset, the dataset needs some adjustments to balance. To balance the dataset, firstly, bioactive molecules having the value of pIC50 greater than or equal to 4.5 are used, secondly, the bioactive molecules belonging to active are twice. These adjustments made the data balanced in both active and inactive classes. However, the range of pIC50 values of bioactive molecules belongs to the intermediate class and fits in both classes active and inactive classes.

Now the dataset is balanced and consisted of 76 bioactive molecules. This balanced dataset is used to train and test the model.

To assess the performance of regression models, a pair of statistical variables: Pearson's correlation coefficient ($R^2$) and root mean square error (RMSE) is used. The $R^2$ value indicates the degree of relationship between two variables (the target and the ground truth). It ranges from -1 to +1 where a positive value shows a positive correlation between the two variables and a negative value shows a negative correlation [33]. RMSE represents the relative error of the predictive model. Finally, a comparative analysis of regression models is carried out.

### 3.3. Module C: ADMET Analysis

The prediction of Absorption, Distribution, Metabolism, Excretion, and Toxicity (ADMET) properties plays a significant role in the drug design process. To evaluate the pharmacokinetics, medicinal chemistry, lipophilicity, water solubility, physicochemical characteristics, and drug-likeness of bioactive compounds; ADMET analysis is performed through the SWISSADME platform [35]. 3D structure of predictive bioactive compounds is retrieved from PubChem using QSAR modeling. The description related to CHEMBL ID, Molecular Formula, PubChem ID, Isomeric SMILES, and 3D Structure of these bioactive molecules are given in Table 3. The structure of these chemical compounds is converted into SMILES and fed to SWISSADME webserver for ADMET analysis. The result of the ADMET analysis decides whether a compound can be a potential drug-like candidate or not. This would help to filter the bioactive molecules for further analysis.

### 3.4. Module D: Molecular Docking

The crystal structure of 3C-like protease (3CLpro) (*PDB ID: 7JSU*) is fetched from the RCSB Protein Data Bank website. This structure is purified by removing ligand, water molecules, and alternative side chains. The protein is prepared by adding polar hydrogen atoms and distributing Kollman charges. After making this macromolecule in charged form, set up a GridBox to cover the active side of the *7JSU* protein. The dimensions of x, y, and z are 30, 30, and 30 with spacing 1, and the centers of x, y, and z are -11.046, 12.826, and 67.789 respectively. Auto-Dock VINA [36] with default parameters is used to prepare protein and ligand to perform molecular docking. The bioactive molecules that are identified after ADMET analysis are used as ligands in molecular docking. After preparing the ligands, molecular docking is performed to calculate the binding affinities (kcal/mol) of these ligands with target protein *7JSU*. The interaction with the lowest binding energy is the best pose. Fig.4 depicts the crystal structure of SARS-CoV-2 3CL protease *7JSU* with a resolution of 1.83 Å.

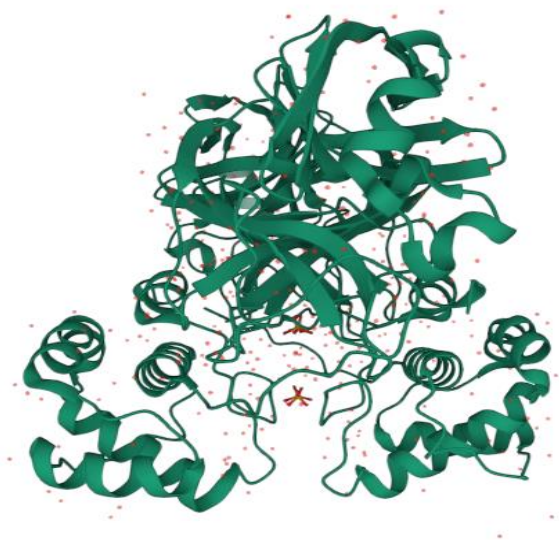

Fig.4. Crystal structure of SARS 3CL protease *7JSU*

## 4. Results and Discussion

In this section, we will discuss the exploratory data analysis, evaluation of the proposed model, comparative analysis, ADMET analysis, and molecular docking, respectively.

### 4.1. Exploratory Data Analysis

Lipinski's rule-of-five descriptor is used to perform exploratory data analysis. The chemical space of the descriptor shows the structure-activity relationship. The bioactive compounds were categorized as active, inactive, and intermediate classes based on their IC50 value as described earlier.

Statistical analysis is used to find the significant difference between both active and inactive classes. For this purpose, the Mann-Whitney U test is employed. It is nonparametric and used to determine whether the dependent variable differs between two independent groups. It evaluates if the dependent variable's distribution is the same for the two groups and, consequently, comes from the same population. Table 2. illustrates the Mann-Whitney U test results regarding significant differences in both bioactivity classes.

Table 2. Mann-Whitney U test Results

| **Descriptor** | **Statistics** | **P** | **Alpha** | **Interpretation** |
|---|---|---|---|---|
| LogP | 440 | 0.4892 | 0.05 | Same distribution (fail to reject H0) |
| MW | 232 | 0.0023 | 0.05 | Different distribution (reject H0) |
| NumHAcceptors | 214.5 | 0.0009 | 0.05 | Different distribution (reject H0) |

| NumHDonors | 157 | 0.00002 | 0.05 | Different distribution (reject H0) |
| pIC$_{50}$ | 0 | 1.37E-09 | 0.05 | Different distribution (reject H0) |

This table indicates that both classes are different. The interpretation of four descriptors MW, NumHacceptors, NumHDonors, and pIC50 highlights that both classes are significantly different except logP descriptor. This descriptor shows no significant difference between the two classes. LogP values for the active class are the lowest, while; the other class' differences are minuscule. Hence, for logP, the test gave no significant difference.

Fig.5. (a-e) represents the box plots of Mann-Whitney U test results in which LogP, MW, NumHAcceptors, NumHDonors, and pIC50 are shown in 5a, 5b, 5c, 5d, and 5e, respectively. LogP is a commonly utilized metric for figuring out a compound's lipophilicity as well as its permeability and penetration of membranes. However, the molecular weight (MW) of a substance is very important to estimate the right size of a compound. Its numerical values are crucial for transit through a lipid membrane. On the other hand, NumHAcceptors, and NumHDonors are used to measure the hydrogen bonding capacity and refer to the quantity of hydrogen bond acceptors and donors, respectively.

According to an analysis of the box plots, the margins of the boxes indicated that there was a negligible difference between both bioactivity classes for LogP. However, MW, NumHAcceptors, NumHDonors, and pIC50 box plots revealed a clear difference between both classes.

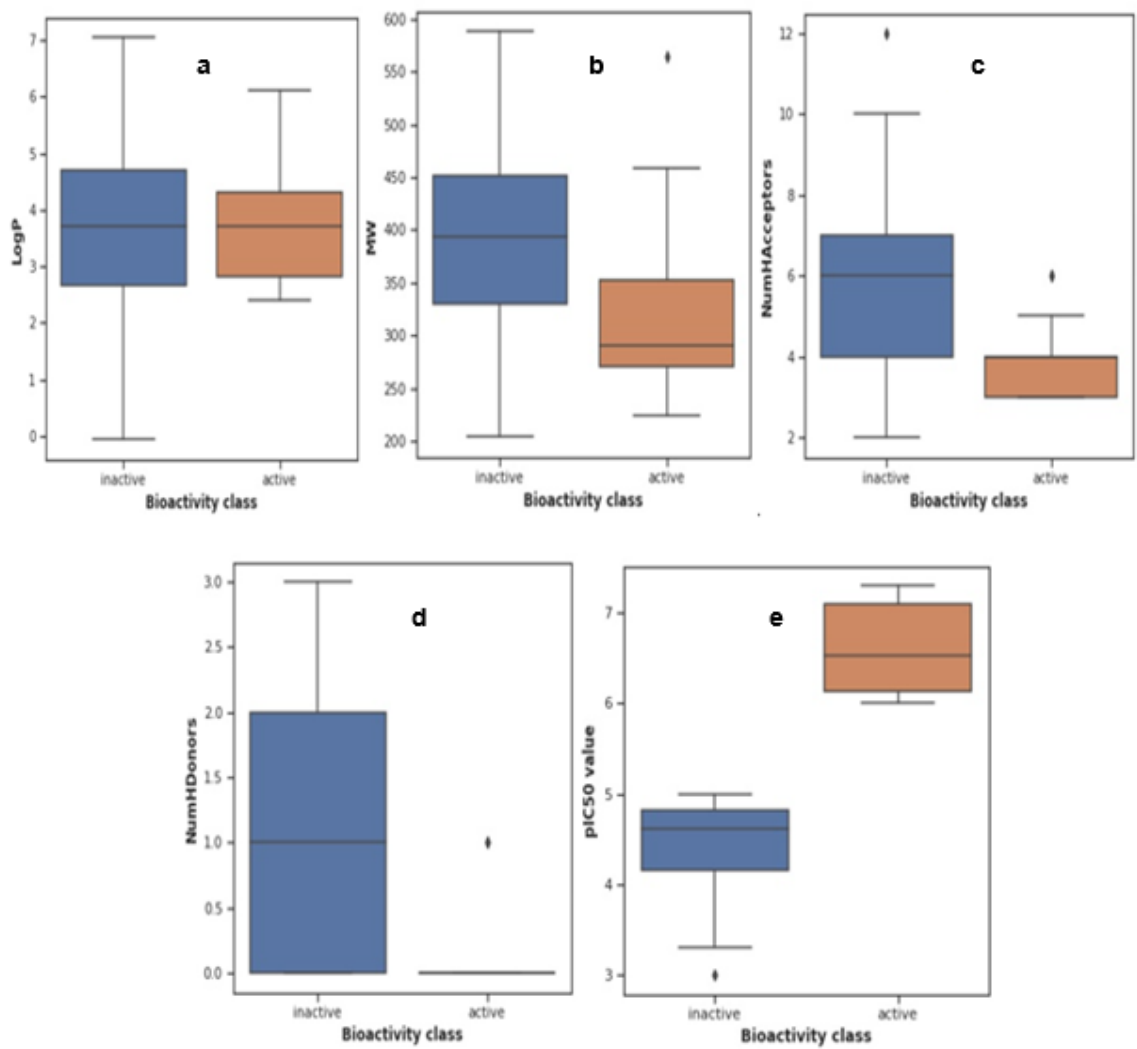

Fig.5 (a-e): Mann-Whitney U test of LogP, MW, NumHacceptors, NumHdonors, and $pIC_{50}$.

From this EDA analysis, we found fifteen bioactive molecules. These fifteen bioactive molecules with their ChEMBL IDs, chemical formulae, PubChem IDs, Isomeric SMILES, and 3D structures are tabulated in Table 3. For further analysis, these bioactive molecules belonging to the active class are selected.

Table 3. Description of Fifteen Bioactive Molecules

| CHEMBL ID | Molecular Formula | PubChem ID | Isomeric SMILES | 3D Structure |
|---|---|---|---|---|
| CHEMBL 187460 | $C_{19}H_{20}O_3$ | 160254 | C[C@H]1COC2=C1C(=O)C(=O)C3=C2C=CC4=C3CCCC4(C)C | |

| CHEMBL 190743 | $C_{17}H_{10}INO_2S$ | 11796320 | C1=CC=C2C(=C1)C=C(S2)CN3C4=C(C=C(C=C4)I)C(=O)C3=O |
| --- | --- | --- | --- |
| CHEMBL 212218 | $C_{14}H_7Cl_2F_3N_2O_6S$ | 2799606 | CC1=CC(=C(C=C1Cl)Cl)S(=O)(=O)C2=C(C=C(C=C2[N+](=O)[O-])C(F)(F)F)[N+](=O) [O-] |
| CHEMBL 212454 | $C_{18}H_8Cl_6O_6S$ | 2774892 | C1=CC(=CC=C1OC(=O)C(=C(Cl)Cl)Cl)S(=O)(=O)C2=CC=C(C=C2)OC(=O)C(=C(Cl)Cl)Cl |
| CHEMBL 222234 | $C_{10}H_6BrNO_3$ | 16203681 | C1=COC(=C1)C(=O)OC2=CC(=CN=C2)Br |
| CHEMBL 222628 | $C_9H_5ClN_2O_2S$ | 16203796 | C1=C(C=NC=C1Cl)OC(=O)C2=CSC=N2 |
| CHEMBL 222735 | $C_{13}H_{10}ClNO_3$ | 16204324 | COC1=CC=CC(=C1)C(=O)OC2=CC(=CN=C2)Cl |
| CHEMBL 222769 | $C_{16}H_9Cl_2NO_3$ | 16203797 | C1=CC(=CC=C1C2=CC=C(O2)C(=O)OC3=CC(=CN=C3)Cl)Cl |
| CHEMBL 222840 | $C_{10}H_6ClNO_3$ | 7230550 | C1=COC(=C1)C(=O)OC2=CC(=CN=C2)Cl |
| CHEMBL 222893 | $C_{14}H_8ClNO_2S$ | 2800273 | C1=CC=C2C(=C1)C=C(S2)C(=O)OC3=CC(=CN=C3)Cl |

| CHEMBL 225515 | $C_{14}H_9ClN_2O_2$ | 16204322 | C1=CC=C2C(=C1)C=C(N2)C(=O)OC3=CC(=CN=C3)Cl | 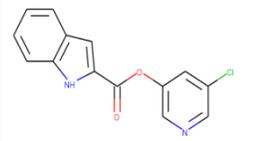 |
|---|---|---|---|---|
| CHEMBL 358279 | $C_{20}H_{14}N_2O_3$ | 515964 | C1=CC=C2C=C(C=CC2=C1)CN3C4=C(C=C(C=C4)C(=O)N)C(=O)C3=O | 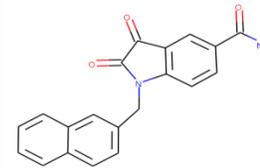 |
| CHEMBL 363535 | $C_{18}H_{12}O_3$ | 114917 | CC1=C2C=CC3=C(C2=CC=C1)C(=O)C(=O)C4=C3OC=C4C | 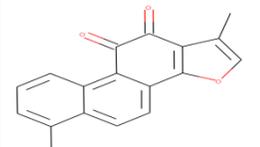 |
| CHEMBL 365134 | $C_{17}H_{10}BrNO_2S$ | 11667869 | C1=CC=C2C(=C1)C=C(S2)CN3C4=C(C=CC=C4Br)C(=O)C3=O | 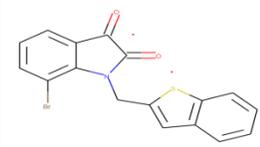 |
| CHEMBL 426898 | $C_{14}H_8ClNO_3$ | 16204318 | C1=CC=C2C(=C1)C=C(O2)C(=O)OC3=CC(=CN=C3)Cl | 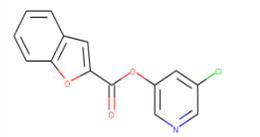 |

### 4.2. Evaluation of proposed model

The proposed QSAR model is developed using the ETR algorithm for 76 bioactive molecules in the balanced dataset. For this purpose, X and Y data matrices are prepared in which 881 PubChem fingerprints are placed in the X matrix and their corresponding pIC50 values are placed in the Y matrix. To train the model, the input dataset is split into a 70/30 training to testing ratio. Here, 70 represents the dataset used to train the model, and 30 represents the dataset used to test the model. Now the QSAR Model is evaluated using two well-known performance measures, Pearson's correlation coefficient ($R^2$) and root mean square error (RMSE). The $R^2$ value indicates the degree of correlations between experimental and predicted values of pIC50. It ranges from -1 to +1 in which positive values show a positive correlation between two variables and negative values highlight the negative correlation [33]. On the other hand, 0.0 shows no relationship; -0.1 negative correlation or +0.1 positive correlation shows the weak relationship; -0.3 (for negative correlation) or +0.3 (for positive correlation) shows the moderate relationship; -0.5 (for negative correlation) or +0.5 (for positive correlation) shows the strong relationship; and -1.0 (for negative correlation) or +1.0 (for positive correlation) shows the perfect relationship. However, RMSE represents the relative error between experimental and model predicted values of pIC50. Fig.6. demonstrates the correlation between the experimental and model predicted pIC50 for training and testing data.

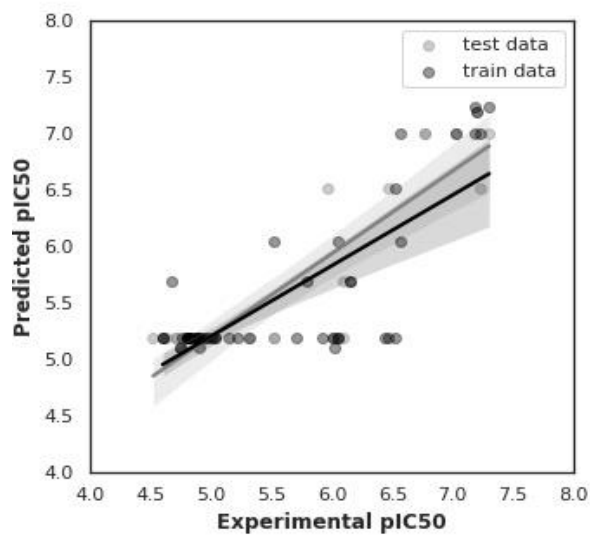

Fig.6. Correlation between the experimental and ETR model predicted pIC50 for training and testing data

Most commonly, in QSAR modeling, the performance is measured in terms of the difference between the values of $R^2$ and $Q^2$. This difference should be less than 0.3.[37] Moreover, the value of $Q^2$ greater than 0.5 shows a good regression performance of the model, and above 0.9 shows excellent performance. Our model has obtained an $R^2$ value of 0.63 for training data and a $Q^2$ value of 0.73 for testing data. There is a minor difference (0.10) between $R^2$ and $Q^2$. This indicates that the proposed QSAR prediction model is the most suitable having sufficient estimation power of pIC50.

### 4.3. Comparative analysis

The performance of the proposed QSAR model is compared with five other state-of-the-art models. These prediction models are trained and their performances are evaluated using two performance measures of Pearson Correlation Coefficient ($R^2$) and RMSE for testing the dataset. The graphical performance comparison among various regression models in terms of experimental and predicted pIC50 values is shown in Fig.7. (a-f). The highest value of $R^2$ =0.73, for testing data, highlights a strong relationship between the experimental and predicted pIC50 values. Fig.7.b shows the relationship between experimental pIC50 and predicted pIC50 values of GradientBoosting Regressor (GBR) and the corresponding value of $R^2$ is 0.62. However, Fig.7.c shows $R^2$ value 0.59 for XGBoost Regressor (XGBR). Fig.7 (d-f) shows a relatively lower correlation of 0.59, 0.58, and 0.52 for Support Vector Regressor (SVR), Decision Tree Repressor (DTR), and Random Forest Repressor (RFR), respectively.

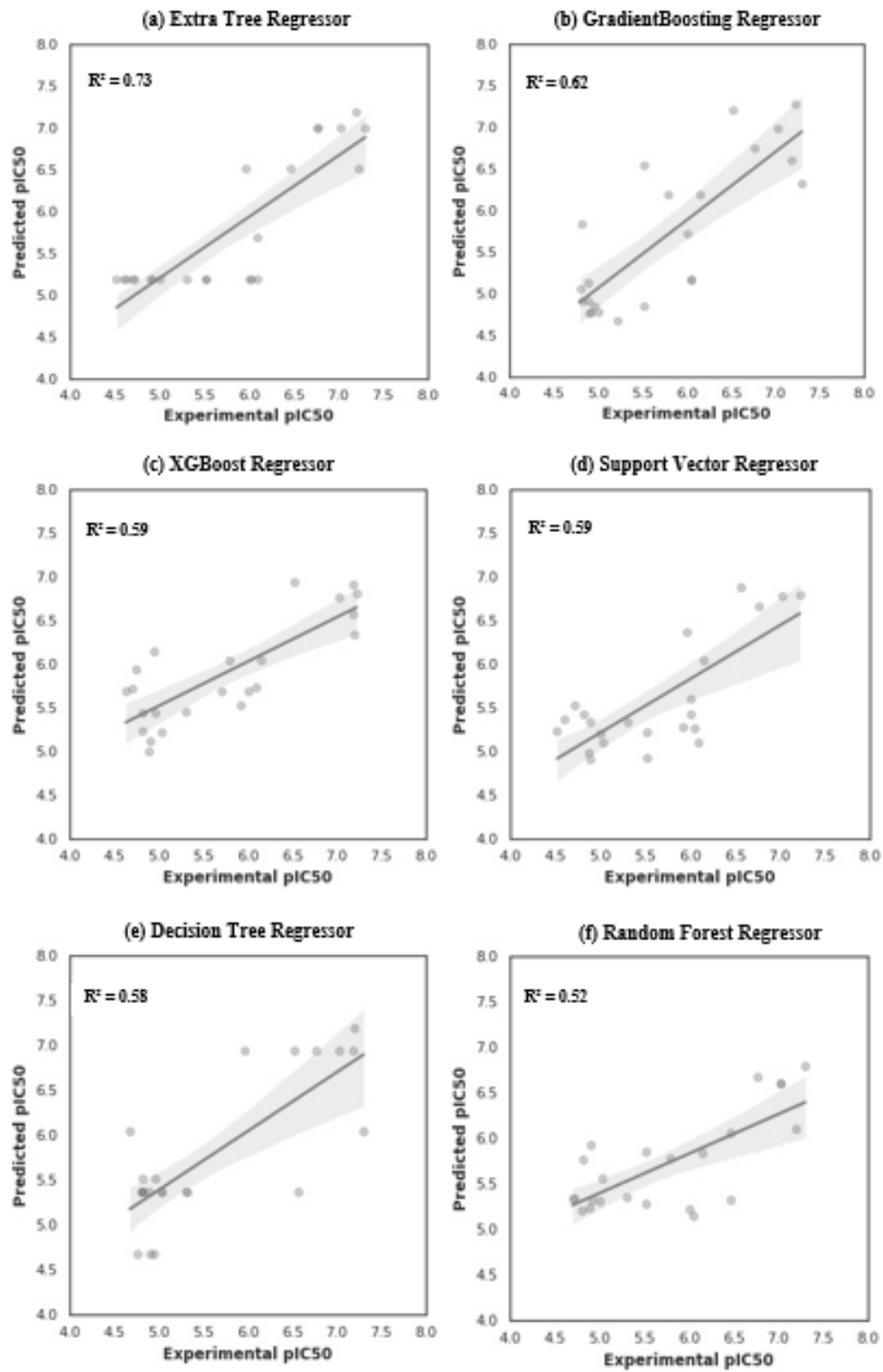

Fig.7 (a-f). Correlation between the experimental and predicted pIC50

Table 4 highlights the performance comparison of various regression models in terms of $R^2$, Mean Squared Error (MSE), and RMSE values. A smaller RMSE value indicates that the model predicts the data accurately. This table indicates the lowest RMSE value of 0.074 in our ETR model. GBR model has a higher RMSE value of 0.078 than our proposed model. Similarly, the higher values of RMSE 0.089, 0.078, 0.092, and 0.089 for XGBR, SVR, DTR, and RFR, respectively. Fig.8.depicts the bar chart of various state-of-the-art regression models in terms of $R^2$ and RMSE.

Table 4. Performance Comparison of various Regression Models

| Regression Model | R-Squared | MSE | RMSE |
|---|---|---|---|
| Extra Tree Regressor | 0.73 | 0.005 | 0.074 |
| Gradient Boosting Regressor | 0.62 | 0.006 | 0.078 |
| XGBoost Regressor | 0.59 | 0.008 | 0.089 |
| Support Vector Regressor | 0.59 | 0.006 | 0.078 |
| Decision Tree Regressor | 0.58 | 0.008 | 0.092 |
| Random Forest Regressor | 0.52 | 0.008 | 0.089 |

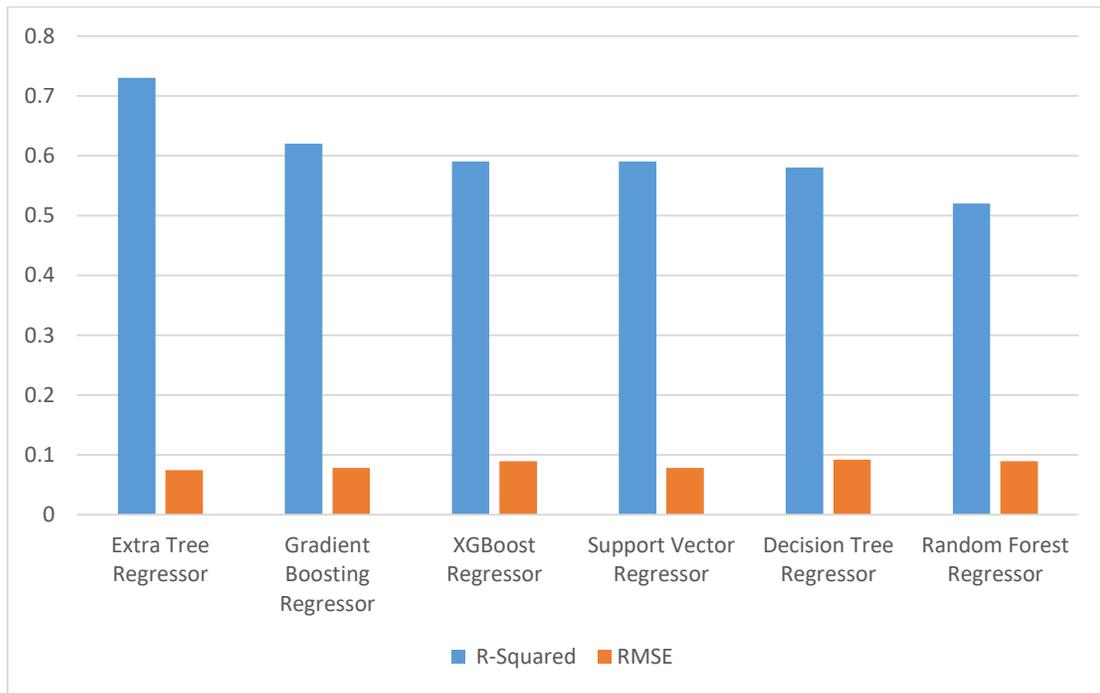

Fig.8. Bar chart performance comparison of Regression Models

### 4.4. ADMET Analysis

The ADMET analysis of the bioactive molecules belonging to the active class is carried out in terms of six main categories such as physicochemical properties, lipophilicity, water solubility, pharmacokinetics, drug-likeness, and medicinal chemistry. There are ten measures in physicochemical properties, five measures in lipophilicity, three measures in water solubility, nine in pharmacokinetics, six in drug-likeness, and four measures in medicinal chemistry. In the study for ADMET analysis, we have computed a total of 37 properties related to absorption, distribution, metabolism, excretion, and toxicity.

The physicochemical properties are expressed in terms of MW, Topological Polar Surface Area (TSPA), Number of Hydrogenbonds Acceptor (NHA), and Number of Hydrogenbonds Donors (NHD). NHA and NHD relevant to the polarity are the main measures of the drug-like molecules. The drug needs to be relatively non-polar to pass through most membranes. A drug needs to be polar to be water-soluble. Overly nonpolar drugs may not be water-soluble, or they may attach to dietary ingredients or blood proteins too firmly. These values related to physicochemical properties are shown in column 2 of Table 5. Further, the values of lipophilicity are calculated in terms of consensus log $P_{o/w}$. Its value is computed by taking an average of values of ilogp, xlogp3, wlogp, mlogp, and logp (silicos-IT). These values are given in column 3 of Table 5. The water solubility is categorized into five classes named as poorly soluble, soluble, moderately soluble, highly soluble, and very soluble. These values are given in column 4 of Table 5.

The Pharmacokinetics properties are expressed in terms of the three most significant measures Gastrointestinal (GI) absorption, Blood Brain Barrier (BBB) permeant, and skin permeation (log Kp). The high value of GI absorption means the drug is absorbable and the low means not absorbable. However, the binary value (yes) of the BBB measure indicates that bioactive molecules can penetrate from BBB and the binary value (no) shows that they cannot penetrate from BBB. On the other hand, the value of Skin permeability (Kp) serves as a proxy for the number of molecules absorbed in the skin; the more negative value indicates the lesser skin absorption. These values related to pharmacokinetics properties are computed and given in column 5 in Table 5.

The binary values of Drug-likeness properties are expressed as yes/no. Its yes value is computed, if 3 of 5 rules are satisfied according to drug-likeness criteria such as Lipinski, Ghose, Veber, Egan, and Muegge rules otherwise, it is assigned no. These values are given in column 6 of Table 5. The property of medicinal chemistry is measured in terms of synthetic accessibility. Its numerical score for drug-like molecules has in the range of 1-10. Numerical value one indicates that it is very easy to synthesize and ten indicates that it is very difficult to synthesize for chemist guidance. This numerical value is given in column 7 of Table 5.

Table 5. ADMET Analysis

| ChEMBL ID | Physicochemical Properties | Lipophilicity | Water Solubility | Pharmacokinetics | Drug-likeness | Medicinal Chemistry |
|---|---|---|---|---|---|---|
| 187460 | MW 324.58 g/mol<br>TSPA 43.37 Å²<br>NHA=3<br>NHD=0 | Consensus log $P_{o/w}$ 3.06 | Moderately soluble | GI absorption=High<br>BBB Permeant=Yes<br>Skin Permeation (log $K_p$)= -5.57cm/s | Yes | Synthetic accessibility = 5.25 |
| 190743 | MW 442.42 g/mol<br>TSPA 69.47 Å²<br>NHA=2<br>NHD=0 | Consensus log $P_{o/w}$ 2.43 | Moderately soluble | GI absorption=High<br>BBB Permeant=Yes<br>Skin Permeation (log $K_p$)= -6.91cm/s | Yes | Synthetic accessibility = 5.36 |
| 212218 | MW 459.18 g/mol<br>TSPA 136.16 Å²<br>NHA=9<br>NHD=0 | Consensus log $P_{o/w}$ 3.58 | Moderately soluble | GI absorption=Low<br>BBB Permeant=No<br>Skin Permeation (log $K_p$)= -5.64cm/s | Yes | Synthetic accessibility = 3.13 |
| 212454 | MW 585.19 g/mol<br>TSPA 86.04 Å²<br>NHA=6<br>NHD=0 | Consensus log $P_{o/w}$ 3.73 | Poorly soluble | GI absorption=Low<br>BBB Permeant=No<br>Skin Permeation (log $K_p$)= -6.33cm/s | No | Synthetic accessibility = 8.41 |
| 222234Y | MW 276.13 g/mol<br>TSPA 51.21 Å²<br>NHA=4<br>NHD=0 | Consensus log $P_{o/w}$ -0.82 | Highly soluble | GI absorption=High<br>BBB Permeant=No<br>Skin Permeation (log $K_p$)= -10.08 cm/s | Yes | Synthetic accessibility = 5.02 |
| 222628 | MW 246.71 g/mol<br>TSPA 59.44 Å²<br>NHA=4<br>NHD=0 | Consensus log $P_{o/w}$ -1.05 | Highly soluble | GI absorption=High<br>BBB Permeant=No<br>Skin Permeation (log $K_p$)= -9.62 cm/s | Yes | Synthetic accessibility = 4.63 |
| 222735Y | MW 280.81 g/mol<br>TSPA 35.53 Å²<br>NHA=4<br>NHD=0 | Consensus log $P_{o/w}$ 0.73 | Very soluble | GI absorption=High<br>BBB Permeant=Yes<br>Skin Permeation (log $K_p$)= -8.19cm/s | Yes | Synthetic accessibility = 5.74 |
| 222769 | MW 350.28 g/mol<br>TSPA 43.37 Å²<br>NHA=4<br>NHD=0 | Consensus log $P_{o/w}$ 0.58 | Very soluble | GI absorption=High<br>BBB Permeant=Yes<br>Skin Permeation<br>(log $K_p$)= -9.57cm/s | Yes | Synthetic accessibility = 6.00 |
| 222840 | MW 231.68 g/mol<br>TSPA 51.21 Å²<br>NHA=4<br>NHD=0 | Consensus log $P_{o/w}$ -0.90 | Highly soluble | GI absorption=High<br>BBB Permeant=No<br>Skin Permeation<br>(log $K_p$)= -9.85cm/s | Yes | Synthetic accessibility = 4.74 |
| 222893Y | MW 305.86 g/mol<br>TSPA 58.39 Å²<br>NHA=3<br>NHD=0 | Consensus log $P_{o/w}$ 1.04 | Very soluble | GI absorption=High<br>BBB Permeant=Yes<br>Skin Permeation<br>(log $K_p$)= -7.90cm/s | Yes | Synthetic accessibility = 5.75 |
| 225515Y | MW 287.81 g/mol<br>TSPA 26.30 Å²<br>NHA=4<br>NHD=0 | Consensus log $P_{o/w}$ 0.06 | Very soluble | GI absorption=High<br>BBB Permeant=No<br>Skin Permeation<br>(log $K_p$)= -9.53cm/s | Yes | Synthetic accessibility = 5.50 |
| 358279 | MW 360.57 g/mol<br>TSPA 80.47 Å²<br>NHA=3<br>NHD=1 | Consensus log $P_{o/w}$ 2.06 | Soluble | GI absorption=High<br>BBB Permeant=No<br>Skin Permeation<br>(log $K_p$)= -6.45cm/s | Yes | Synthetic accessibility = 4.55 |
| 363535Y | MW 301.48 g/mol<br>TSPA 51.21 Å²<br>NHA=3<br>NHD=0 | Consensus log $P_{o/w}$ 2.29 | Soluble | GI absorption=High<br>BBB Permeant=Yes<br>Skin Permeation<br>(log $K_p$)= -5.95cm/s | Yes | Synthetic accessibility = 5.35 |
| 365134 | MW 393.40 g/mol<br>TSPA 69.47 Å²<br>NHA=2<br>NHD=0 | Consensus log $P_{o/w}$ 2.05 | Soluble | GI absorption=High<br>BBB Permeant=Yes<br>Skin Permeation<br>(log $K_p$)= -7.30cm/s | Yes | Synthetic accessibility = 6.47 |
| 426898Y | MW 289.80 g/mol<br>TSPA 43.37 Å²<br>NHA=4<br>NHD=0 | Consensus log $P_{o/w}$ 0.57 | Very soluble | GI absorption=High<br>BBB Permeant=Yes<br>Skin Permeation<br>(log $K_p$)= -8.23cm/s | Yes | Synthetic accessibility = 5.60 |

The most important characteristics, which have been taken into account in most of the metrics used to create limits in the drug-like chemical space, are lipophilicity, molecular size, and polarity [38]. There is substantial proof that drugs with higher lipophilicity and molecular weight, such as those with high molecular corpulence, are more likely to be dropped during clinical trials. These are linked to complications with oral absorption. For a drug-like compound, the numerical values of MW should be less than 480g/mol. This table shows that ChEMBL ID 212454 has a relatively higher value of MW 585.19g/mol than the stated criteria. All physicochemical properties of this compound such as TSPA, NHA, and NHD are satisfied except for the numerical value of MW. Since MW measure is the most important physicochemical property. Further, this compound has also low GI absorption, is poorly soluble and violates the drug-likeness rules 3 out of 5 drug-likeness criteria. Therefore, this compound cannot be a potential drug candidate.

On the other hand, ChEMBL ID 212218 has the value of TSPA 136.16 Å². That is greater than the standard criterion TSPA value of 130 Å². It cannot penetrate the blood-brain barrier and it has poor penetration through the cell membrane. Further, this compound has also low GI absorption. Therefore, this compound does not satisfy pharmacokinetic properties as well. We infer that this compound cannot be a potential drug candidate.

Other thirteen compounds that fulfill the ADMET criteria, can be potential drug candidates. For further validation, these thirteen compounds were investigated for the molecular docking process.

### 4.5. Molecular Docking

The filtered molecules are obtained from ADMET analyses that belong to the active class. In the final stage of the current study, molecular docking is performed, and check the binding affinity of the selected inhibitors with the target protein *7JSU*. The value of binding affinity is represented in the unit of Kcal/mol. In the ligand-based docking approach, we want to investigate the efficacy of the selected inhibitors. Fig.9. visually depicts the best pose of six bioactive molecules towards the target protein that corresponds to the lowest binding affinity in the range of -7.0 to -8.4. The best ligands pose towards the target protein corresponds to the most negative value binding energy.

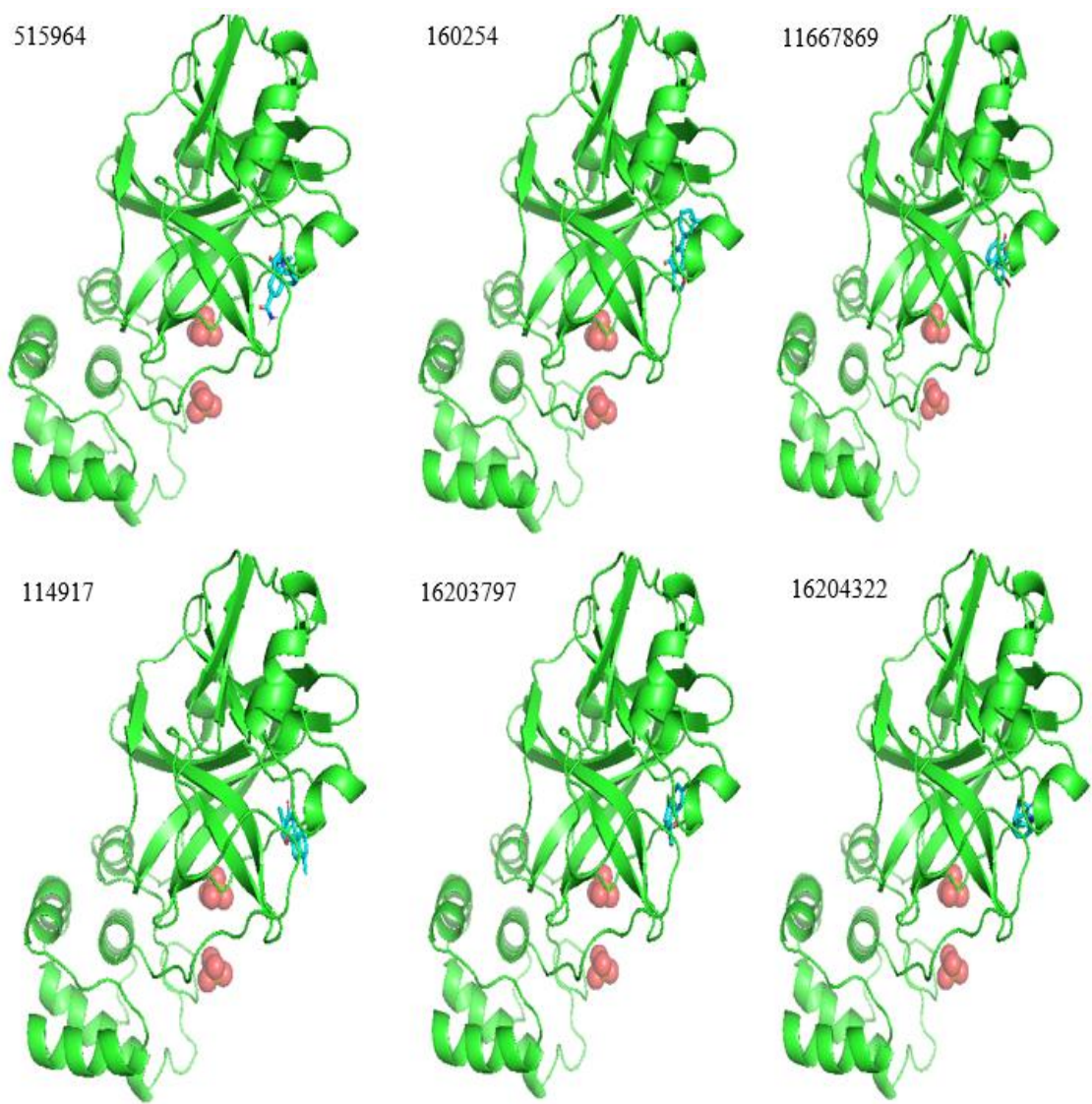

Fig.9. Best pose of six bioactive molecules towards target protein *7JSU*.

Table 6. demonstrated the results of binding affinities and RMSD of thirteen selected ligands computed molecular docking towards the target protein *7JSU*. The lower RMSD measure indicates the higher correctness in the docking geometry of the ligand molecule from its reference position in the original protein complex. In the study, corresponding to the best pose, we found all thirteen selected ligands have zero RMSD value.

In general, a lower value of binding affinity means a stronger interaction between protein and ligand. From the results in Table 6, we have selected six bioactive molecules with ChEMBL ID 187460, 222769, 225515, 358279, 363535, and 365134, that possess the lowest binding affinity in the range of -7.0 to -8.4. So these can be the potential drug candidates. The molecular docking result suggested that the ChEMBL ID 358279 is the most suitable drug candidate among these six candidates with the lowest binding affinity of -8.4. It means this compound has the strongest interaction with the target protein *7JSU* than other bioactive molecules.

Table 6. Docking results of thirteen selected ligands towards the target protein *7JSU*

| Protein Name | ChEMBL ID | Ligand ID | Best Pose | Binding Affinity (kcal/mol) | Distance from Best Mode | |
| --- | --- | --- | --- | --- | --- | --- |
| | | | | | RMSD lower bond | RMSD Upper bound |
| 7JSU | 187460 | 160254 | 1 | -8.0 | 0.000 | 0.000 |
| 7JSU | 190743 | 11796320 | 1 | -6.7 | 0.000 | 0.000 |
| 7JSU | 222234 | 16203681 | 1 | -5.4 | 0.000 | 0.000 |
| 7JSU | 222628 | 16203796 | 1 | -5.4 | 0.000 | 0.000 |
| 7JSU | 222735 | 16204324 | 1 | -6.6 | 0.000 | 0.000 |
| 7JSU | 222769 | 16203797 | 1 | -7.3 | 0.000 | 0.000 |
| 7JSU | 222840 | 7230550 | 1 | -5.3 | 0.000 | 0.000 |
| 7JSU | 222893 | 2800273 | 1 | -6.5 | 0.000 | 0.000 |
| 7JSU | 225515 | 16204322 | 1 | -7.0 | 0.000 | 0.000 |
| 7JSU | 358279 | 515964 | 1 | -8.4 | 0.000 | 0.000 |
| 7JSU | 363535 | 114917 | 1 | -7.6 | 0.000 | 0.000 |
| 7JSU | 365134 | 11667869 | 1 | -7.8 | 0.000 | 0.000 |
| 7JSU | 426898 | 16204318 | 1 | -6.6 | 0.000 | 0.000 |

## 5. Conclusion

In this study, we repurposed existing therapeutic agents by examining drug-like bioactive molecules for Covid-19. We developed a hybrid approach that combines useful extracted information through various bioinformatics tools. The Main protease, 3C-like protease (3CL$^{pro}$) is a most suitable potential drug target among coronavirus proteins due to its property as a viral enzyme. The current in-silico study has explored the small molecule inhibitors against the infection of COVID-19. The virtual screening is performed on the ChEMBL dataset and found 133 bioactive molecules against 3CL$^{pro}$. QSAR modeling is developed to predict the chemical compounds having the best biological activities.

Our comparative analysis demonstrated that the proposed Extra Tree Regressor (ETR) based QSAR model has improved prediction results related to the bioactivity of chemical compounds as compared to Gradient Boosting, XGBoost, Support Vector, Decision Tree, and Random Forest based regressor models.

From Lipinski's rules, we found 133 drug-likeness bioactive molecules against SARS coronavirus 3CL Protease. From ADMET analysis on active class data, we identified thirteen novel bioactive molecules for SARS-Cov-2. In the next step, the efficacy of bioactive molecules is computed in terms of binding affinity using molecular docking. This technique has shortlisted the six most suitable bioactive molecules with ChEMBL IDs 187460, 222769, 225515, 358279, 363535, and 365134. These molecules can further be investigated as drug candidates for SARS-COV-2 3CL Protease. The pharmacologist community can adopt these short-listed relatively small-size bioactive molecules to develop potential drug candidates.